\documentclass[prd,superscriptaddress,nofootinbib,floatfix, 11pt]{revtex4-1}
\usepackage[colorlinks=true,
linkcolor=blue,
urlcolor=magenta,
citecolor=blue]{hyperref}
\usepackage[utf8]{inputenc} 
\usepackage{xcolor}
\usepackage{multirow}
\usepackage[a4paper, left=2.5cm, right=2.5cm, top=2.5cm, bottom=2.5cm]{geometry}
\usepackage[T1]{fontenc}
\usepackage{amsmath,amssymb,amsthm}  
\usepackage{amsfonts}
\usepackage{mathrsfs}      
\usepackage{latexsym,bm}
\usepackage{graphicx}      
\usepackage{wrapfig}    
\usepackage{indentfirst}
\usepackage{slashed}  
\usepackage{units} 
\usepackage{fancybox}       
\usepackage{array}    
\usepackage{tabularx}    
\usepackage{booktabs}    
\usepackage{subcaption} 
\usepackage{makecell}
\usepackage{caption}
\usepackage{ulem}
\usepackage{rotating} 
\usepackage{booktabs} 
\usepackage{cancel}

\newcommand{\beq}{\begin{eqnarray}}
\newcommand{\eeq}{\end{eqnarray}}
\newcommand{\ben}{\begin{enumerate}}
\newcommand{\een}{\end{enumerate}}

\allowdisplaybreaks[4]

\begin{document}


\title{Polarization fractions and  helicity-dependent CP asymmetries in $B_{(s)} \to \rho\rho, \rho K^\ast$ and $K^\ast K^\ast$ decays}

\author{Jian Chai} 
\affiliation{School of Physics and Advanced Energy, Henan University of Technology, Zhengzhou 450001, China.}

\author{Shan Cheng}\email{scheng@hnu.edu.cn}
\affiliation{School of Physics and Electronics, Hunan University, 410082 Changsha, China.}
\affiliation{School for Theoretical Physics, Hunan University, 410082 Changsha, China.}

\author{Feng-Qing Hu}
\affiliation{School of Physics and Advanced Energy, Henan University of Technology, Zhengzhou 450001, China.}

\author{Ya Li}\email{liyakelly@163.com }
\affiliation{Department of Physics and Institute of Theoretical Physics, Nanjing Normal University, Nanjing 210023, Jiangsu, China.}

\author{Jin-Yang Shen}
\affiliation{School of Physics and Electronics, Hunan University, 410082 Changsha, China.}

\author{Da-Cheng Yan}\email{yandac@126.com }  
\affiliation{School of Mathematics and Physics, Changzhou University, Changzhou 213164, China.}

\date{\today}

\begin{abstract}

In this paper, we present a phenomenological analysis of $B_{(s)} \to \rho\rho, \rho K^\ast$ and $K^\ast K^\ast$ decays using state-of-the-art perturbative QCD (pQCD) calculations. Our study is primarily motivated by recent polarization measurements from the LHCb and Belle II collaborations, which have significantly improved the precision of polarization fractions and enabled the first full determination of polarization-dependent CP asymmetries. 
This work extends the comprehensive pQCD study of charmless two-body $B$ decays reported in our previous paper [Chin. Phys. C 46 (2022) 123103], with a particular focus on polarization observables, especially the CP asymmetries in each helicity state, which reflect distinct orbital angular momentum configurations between the two vector mesons. Our predictions for the branching ratios and longitudinal polarization fractions in the $B^0 \to K^{\ast 0} {\bar K}^{\ast 0}$ and $B^+ \to \rho^0 K^{\ast +}$ modes are in good agreement with the new experimental data. However, the calculated longitudinal polarization fraction for $B_s \to K^{\ast 0} {\bar K}^{\ast 0}$ is significantly larger than the LHCb measurement. Moreover, the predicted (helicity-dependent) CP asymmetries in $B^+ \to \rho^0 K^{\ast +}$ are about $30 \%$ smaller than the observed values. These discrepancies point to a rich interplay between different topological amplitudes, highlighting the need for further theoretical investigation to resolve the long-standing polarization puzzle in two-body $B$ decays into vector mesons.

\end{abstract}



\maketitle 

\section{Introduction}

Charmless two-body $B$ decays, which receive comparable contributions from tree-level and penguin-loop amplitudes, can exhibit sizable charge-parity (CP) asymmetries \cite{Gambino:2020jvv,Cerri:2018ypt,Kou:2018nap}. Among these, decays into two vector mesons are particularly powerful probes of the underlying weak and QCD dynamics, as the final state can accommodate distinct orbital angular momentum configurations. Beyond the conventional CP-averaged branching fractions and overall CP asymmetries, these modes offer a richer set of observables: the polarization fractions and, more profoundly, the CP asymmetries resolved for each helicity state. 

Under the naive factorization ansatz, the longitudinal polarization fractions of $B \to VV$ decays are expected to be close to unity \cite{Korner:1979ci,Wirbel:1985ji,Bauer:1986bm}. However, experimental measurements span a wide range, from about $10\%$ to nearly $100\%$, 
depending on the specific decay mode, giving rise to the long-standing polarization puzzle. 
This discrepancy suggests that the naive factorization approach is insufficient, motivating extensions with QCD and QED correction effects
\cite{Ali:2007ff,Chen:2002pz,Li:2004ti,Zou:2015iwa,Yan:2018fif,Beneke:2006hg,Cheng:2009cn,Wang:2017rmh,Yan:2025ocu,Li:2021qiw}\footnote{ 
The final-state interactions \cite{Colangelo:2004rd,Cheng:2004ru} have also been proposed as a possible resolution.}.

With increased integrated luminosity, the LHCb collaboration has reported the unprecedented high-precision measurement of (helicity-dependent) CP asymmetries in $B^+ \to \rho^0 K^{\ast +}$ decay \cite{LHCb:2025zvw}. The results indicate that the overall CP asymmetry of squared magnitude ${\cal A}_{\rm CP}$ 
is predominantly driven by its longitudinal component ${\cal A}_{\rm CP}^0$: 
\beq  {\cal A}_{\rm CP}(B^+ \to \rho^0 K^{\ast +}) = (50.7 \pm 6.2 \pm 2.4) \%, \nonumber \\
{\cal A}_{\rm CP}^0(B^+ \to \rho^0 K^{\ast +}) = (66.4 \pm 8.3 \pm 2.9) \%, \eeq 
where the first and second uncertainties are statistical and systematic, respectively.
In addition, the Belle-II collaboration have presented a measurement of the branching fraction, the longitudinal polarization fraction, and the time-dependent CP violation parameters for $B \to \rho^+ \rho^-$ decay \cite{Belle-II:2024frs}. 
Furthermore, the LHCb collaboration has updated the measurements of the longitudinal polarization fractions for the $B^+ \to \rho^0 K^{\ast +}, B_{(s)}^0 \to K^{\ast 0} {\bar K}^{\ast 0}$ decays \cite{LHCb:2025zvw,LHCb:2025ftm}. 
\beq 
&&f_0(B^+ \to \rho^0 K^{\ast +}) = (72.0 \pm 2.8 \pm 0.9) \%, \nonumber \\
&&f_0(B^0 \to K^{\ast 0} {\bar K}^{\ast 0}) = (60.0 \pm 2.2 \pm 1.7) \%, \nonumber \\
&&f_0(B_s^0 \to K^{\ast 0} {\bar K}^{\ast 0}) = (15.9 \pm 1.0 \pm 0.7) \%. \eeq 
A surprisingly small longitudinal fraction is confirmed in the $B_{(s)} \to K^{\ast 0} {\bar K}^{\ast 0}$ decay, 
standing in marked contrast to the expectations from various theoretical approaches. Moreover, the resulting $U$-spin ratio, $f_0^{B^0}/f_0^{B_s} = 3.77^{+0.36}_{-0.31}$, highlights the importance role of nonfactorizable contributions \cite{Choudhury:2026avj}.

In this paper, we perform a perturbative QCD (pQCD) calculation for the $B_{(s)} \to \rho\rho, K^\ast \rho$ and $K^\ast K^\ast$ decays. This work is an extension of our previous comprehensive study of $78$ modes of $B \to PP, PV, VV$ decays \cite{Chai:2022ptk}, which was primarily concerned with branching fractions and overall CP asymmetries. Motivated by the recent LHCb measurements \cite{LHCb:2025zvw,LHCb:2025ftm}, we now concentrate on polarization fractions and helicity-dependent CP asymmetries. Our calculation incorporates all the well-known next-to-leading-order (NLO) corrections, including weak-vertex whose contributions are effectively absorbed into the Wilson coefficients, as well as the quark-loop and chromomagnetic penguin contributions. 
We also include corrections to the longitudinal polarized $B \to V$ form factor. 
In addition, power corrections proportional to the spectator-light-quark momentum fraction $u_1$ and 
the heavy-quark expansion parameter $r_b = m_b/m_B$ are taken into account. 

The paper is organized as follows. In Section \ref{sec:pQCD}, we provide a brief introduction to the pQCD approach for studying $B \to VV$ decays. 
Section \ref{sec:numerics} presents our numerical results and discussions. A summary is given in Section \ref{sec:summary}. 
 
\section{Framework}~\label{sec:pQCD}

The perturbative QCD (pQCD) approach is formulated within the hard-scattering mechanism, which requires a hard gluon exchange between the four-fermion effective vertex and the spectator quark. In this framework, the transverse momentum $k_T$ is picked up to regulate the endpoint singularities that arise from the integration over the longitudinal momentum fraction $u_i$ carried by the anti-quark inside hadrons.

In the weak decays of $B$ mesons, the transverse momentum typically spans three distinct scales: the hard scale ${\cal O}(m_b)$, the hard-collinear scale ${\cal O}(\sqrt{m_b \Lambda})$ and the soft scale $\Lambda$, where $m_b$ denotes the $b$ quark mass and $\Lambda$ represents a typical hadronic scale. In the soft region, the integration over $k_T$ gives rise to large logarithms $\ln^2 (k_T^2/m_b^2)$ and $\ln (k_T^2/m_b^2)$. Since the hard-scattering amplitudes in factorized formulations should not depend on soft dynamics, these large logarithms are naturally resummed into a Sudakov exponent $e^{-s}$. This exponent strongly suppresses contributions from small $k_T$ interaction while enhancing those from the hard-collinear region. Furthermore, an additional double logarithm $\ln^2 u_i$ emerges when the integration over the momentum fraction approaches the on-shell quark propagator. It is furtherly resummed to the threshold factor $S_{\rm th}$ that suppressed the on-shell dynamics, meanwhile, highlight the hard scattering of intermediate propagators.  

\begin{figure}[t] \vspace{-2mm}   
\includegraphics[width=0.6\textwidth]{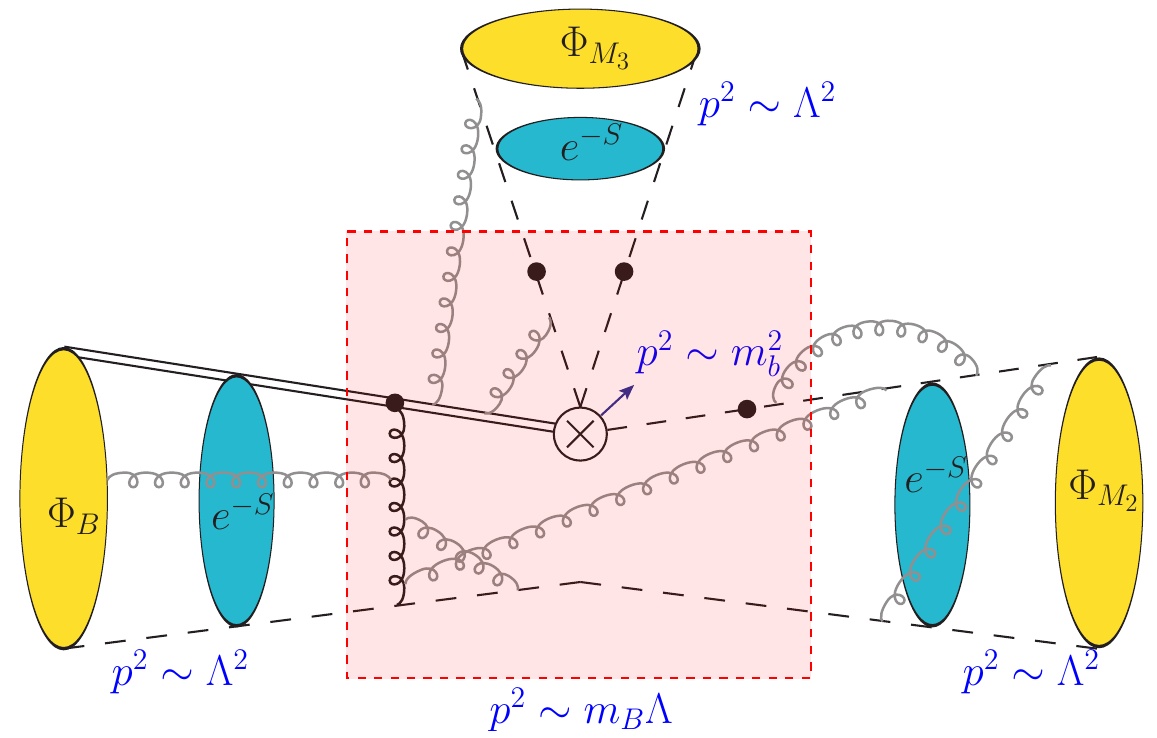} 
\caption{The sketch map of pQCD factorization of $B \to M_2M_3$ decays.} \label{fig:pqcd}
\end{figure}

\begin{figure}[t]
	\begin{center}
		\includegraphics[width=0.9\textwidth]{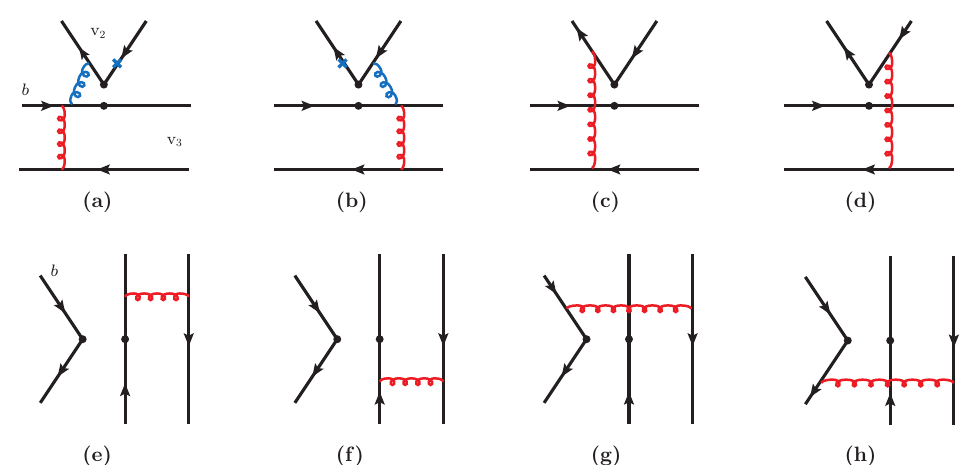}
	\end{center}
	\vspace{-4mm}
	\caption{Typical feynman diagrams contributing to $B_{(s)}\to M_2M_3$ decays at LO. Diagrams (a,b) also depict the NLO vertex correction.} 
	\label{fig:fig2}
\end{figure}

Taken together, these considerations lead to the $k_T$ factorization formulism for the two-body hadronic $B$ decay amplituide, expressed in the convolution of hard kernels and meson LCDAs, as shown in figure \ref{fig:pqcd}. The bullets denote the possible attachments of the hard gluon exchanged between the spectator quark and the four-fermion effective vertex, corresponding to the four types LO diagrams in figure \ref{fig:fig2}. 
\beq &&{\cal M}(B \to M_2M_3) = 
H_{r}(M_W, t_f) \otimes H(t_f,\mu) \otimes \phi(u,P,b,\mu) \nonumber \\
&&= \sum_{i=1}^{10} \sum_{t=2}^{3} \prod_{j=1,2,3} 
C_i(M_W, t_f) \otimes H^t_i(t_f, b) \otimes \phi^{t}_{j}(u,b)  \cdot S_{{\rm th},j}(u)
e^{-s_j(P,b) - \int_{1/b_j}^{t_f} \frac{d\overline{\mu}}{\overline{\mu}} \gamma_{\phi}(\alpha_s(\overline{\mu}))}.
\label{eq:B2MM-fact}\eeq
Here, $H_r$ (with $C_i$) denotes the hard kernel that incorporates the first stage of evolution, stretching from the $m_W$ scale down to the typical hard scale $t_f$. The term $H(t_f, \mu)$ represents the perturbatively calculable hard part, which is encapsulated within the pink box. The convolution $H \otimes \phi $ accounts for the second stage of evolution, from the hard scale $t$ down to $\Lambda_{QCD}$,  where $\phi^t$ refers to the light-cone distribution amplitudes (LCDAs) of the mesons at a given twist. The Sudakov factor $s(P,b)$, with $P$ being the larger components of light-cone momentum for each meson, arises from the resummation of large double logarithms and appears in the exponent. Meanwhile, $\gamma_\phi$ is the anomalous dimension of the wave function, originating from the resummation of single logarithms in the quark self-energy correction. The subscript $i,j$ and $t$ label the contributions from different four-fermion operators, the initial and final mesons, and the various twists of the meson LCDAs, respectively\footnote{Contributions from vertex corrections, quark-loop corrections, and the chromomagnetic operator $O_{8g}$ are absorbed into the corresponding Wilson coefficients.}. 

\begin{figure}[t]
	\begin{center}\vspace{4mm}
		\includegraphics[width=0.9\textwidth]{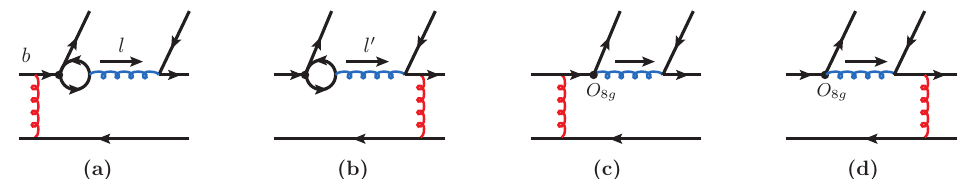}
	\end{center}
	\vspace{-4mm}
	\caption{Quark-loop and Chromomagnetic penguin contributions to $B \to M_2M_3$ decays.}
	\label{fig:fig3}
\end{figure}

At LO, actually ${\cal O}(\alpha_s)$ order due to the hard gluon exchange mechanism, there are eight Feynman diagrams contribute to $B \to M_2M_3$ decays under the pQCD framework, as depicted in figure \ref{fig:fig2}. These diagrams are classified into two categories: emission (a-d) and annihilation (e-h) diagrams. Each category is further divided into factorizable and nonfactorizable contributions, depending on whether the six-fermion interactions can be reduced to a product of form factors and decay constants. 

In the factorizable emission diagrams \ref{fig:fig2}(a, b), we include NLO vertex corrections via the blue gluons, with the additional attachments marked by a cross. Beyond the vertex corrections, whose effects can be absorbed into a redefinition of the Wilson coefficients $C_i$, there are two further NLO corrections associated with the penguin four-fermion operators, saying the quark-loop and chromomagnetic penguin corrections. They produce independent amplitudes corresponding to figures \ref{fig:fig3}(a, b) and (c, d), respectively. The well-known NLO corrections to the $B \to \pi$ transition form factors \cite{Li:2012nk,Cheng:2014fwa} are also taken into account in our calculation of the factorizable emission amplitudes for the longitudinally polarized vector final states. In addition, our analysis includes two extra power corrections, which are proportional to the spectator-quark momentum fraction $u_i$ and the heavy-quark mass ratio $r_b\equiv m_b/m_B$. 

CP asymmetries refer to the differences in the behavior of particles and their antiparticles when both charge and parity are reversed. The presence of at least two decay amplitudes with both a non-zero weak phase difference and a non-zero strong phase difference constitutes the essential conditions for observing CP violation in $B$ meson weak decays. In the perturbative QCD (pQCD) framework, the non-zero weak phase difference originates from the distinct CKM matrix elements. These elements are associated with the effective four-fermion operators constructed at both tree and penguin loop levels.
As for the strong phase difference, there are generally three sources. The first, and also the dominant one, arises from the interplay between emission and annihilation topologies. In this case, the annihilation amplitude can induce a substantial strong phase, whereas the emission amplitude remains predominantly real. The second source comes from the on-shell charm-quark loop shown in figure \ref{fig:fig3}(a,b), which introduces a large imaginary part \cite{Beneke:2000ry,Beneke:2001ev,Beneke:2003zv}. The third, but by no means least important, is the Glauber-gluon correction to the spectator-emission amplitude. In the penguin dominated channels $B \to \rho^0 K^{\ast 0}, K^{\ast 0}{\bar K}^{\ast 0}$ we are interesting here, 
we would not consider the third source of strong phase since it is make sense only for the color-suppressed decays, particularly for $B \to \pi^0\pi^0$ \cite{Liu:2015sra}, by an additional phase to the final-state mesons. 
Additionally, the Sudakov exponent can also yield an imaginary part. However, in practice, this effect only becomes relevant in baryonic decays through the angular distribution of the scattered hadrons.

\section{NUMERICS AND DISCUSSIONS}~\label{sec:numerics}

The detailed expressions for the effective-operator decomposition of these channels can be found in the comprehensive work of Ref. \cite{Chai:2022ptk}. The same reference also provides the complete topological amplitude basis, including the factorizable emission, spectator emission, factorizable annihilation, and spectator annihilation amplitudes, as well as the quark-loop and chromomagnetic-penguin correction functions. The parameter choices for the meson LCDAs are given there as well. We therefore proceed directly to the numerical analysis without quoting these detailed expressions.

In the transversity basis, the decay amplitudes for $B \to VV$ processes can be decomposed into 
the longitudinal ($A_0$), perpendicular ($A_\perp$) and parallel ($A_\parallel$) components 
\beq {\cal M} = \sum_{\lambda = 0, \perp, \parallel} {\cal M}_{\lambda}
= i A_0 + i \left( \epsilon_{2 {\rm T}}^{\ast} \cdot \epsilon_{3 {\rm T}}^{\ast} \right) \frac{A_{\parallel}}{\sqrt{2}} 
+ \left( \varepsilon_{\mu\nu\rho\sigma} \, n_1^\mu n_2^\nu \epsilon_{2 {\rm T}}^{\ast \rho} \epsilon_{3 {\rm T}}^{\ast \sigma} \right) \frac{A_\perp}{\sqrt{2}}.  
\label{eq:amplitude} \eeq
Here $\epsilon^\ast_{j {\rm T}}$ are the transverse polarization vectors of the outgoing vector mesons, satisfying $\epsilon_j^\ast \cdot p_j = 0$ and $\sum_\lambda \epsilon_\lambda \cdot \epsilon^\ast_\lambda = -1$, where $j=2,3$ labels the vector mesons and $\lambda=\parallel, \perp$ denoting the specialized polarizations.
These compoments $A_{0}$, $A_{\parallel}$ and $A_{\perp}$ correspond to amplitudes in which the two vector mesons have relative orbital angular momentum $L =0, 2, 1$, respectively. Their branching fractions are then expressed in terms of the probabilities associated with each polarization amplitude
\beq && {\cal B}_{\lambda} = \frac{\vert \vec{p} \vert}{16 \pi m_{B_{(s)} } } \tau_{B_{(s)}} \left[ \vert {\cal M}_{\lambda}({\bar B}_{(s)} \to f) \vert^2 + 
\vert {\cal M}_{\lambda}(B_{(s)} \to {\bar f}) \vert^2 \right],
\label{eq:Br}\eeq
and the total branching fraction is given by their sum. 
Here, $n_1$ and $n_2$ are unity vectors along the positive and negative directions, respectively, 
$\vec{p} = \vec{p}_2 = \vec{p}_3$ is the 3-momentum of either of the two vector mesons in the final state, 
and $\tau_{B_{(s)}}$ is the lifetime of the $B_{(s)}$ meson. 

Beside the branching fractions, people can define the helicity polarization fractions which average the possibilities over $B \to VV$ and its charge-conjugated channels, as well as the charge-specific quantities by considering only the $B \to VV$ channel or its changed-conjugated channel ${\bar B} \to VV$. 
\beq f_\lambda = \frac{\vert {\bar A}_\lambda \vert^2 + \vert A_\lambda \vert^2}{\sum_{\lambda} 
\left(\vert {\bar A}_\lambda \vert^2 + \vert A_\lambda \vert^2 \right)}, \quad 
f_\lambda^+ = \frac{\vert A_\lambda \vert^2}{\sum_{\lambda} \vert A_\lambda \vert^2}, \quad 
f_\lambda^- = \frac{\vert {\bar A}_\lambda \vert^2}{\sum_{\lambda} \vert {\bar A}_\lambda \vert^2}. \label{eq:observables-2} \eeq
The averaged polarization fractions should satisfy the normalization condition $f_0 + f_{\parallel} + f_{\perp}=1$.  
The charge-specific quantities characterize the polarization content of the particle and antiparticle channels separately and constitute observables of direct CP violation at the level of probabilities. 
As complementary probes of the direct CP violation defined at the amplitude level through the interference of weak and strong phases,
$f_\lambda^+ \neq f_\lambda^-$ provide the corresponding quantitative information in squared moduli. 
In additional, people can define the polarization-resolved asymmetry of squared magnitudes and the associated phase that takes the difference
\beq {\cal A}_{\rm CP}^{\lambda} = \frac{\vert {\bar A}_\lambda \vert^2 - \vert A_\lambda \vert^2}{\vert A_\lambda \vert^2 + \vert {\bar A}_\lambda \vert^2 }, \quad 
\triangle {\cal A}_{\rm CP}^\lambda = {\rm Arg}({\bar A}_\lambda) - {\rm Arg}(A_\lambda). \label{eq:observables-3} \eeq
The direct CP asymmetry ${\cal A}_{\rm CP}$ is then naturally obtained by summing over all polarization, 
\beq {\cal A}_{\rm CP} = \frac{\sum_{\lambda} \left(\vert {\bar A}_\lambda \vert^2 - \vert A_\lambda \vert^2 \right)}{\sum_{\lambda} \left(\vert A_\lambda \vert^2 + \vert {\bar A}_\lambda \vert^2 \right)}. \label{eq:observables-4} \eeq
The large number of the observables make the study of $B \to VV$ decays more opportunity to confirm or specify the phenomenological approaches. 

\begin{table}[th]\begin{center}
\caption{Anatomy of the branching fractions (in unit of $10^{-6}$) from each helicity amplitude.} \label{tab:BRs} 
\setlength{\tabcolsep}{8pt}
\renewcommand{\arraystretch}{1.2}
\begin{tabular}{l  c c c  c c }\hline 
Modes \quad &  \quad ${\cal B}_0$ \qquad & \quad ${\cal B}_{\parallel}$ \quad & \quad ${\cal B}_{\perp}$ \quad & \quad ${\cal B}$ \quad & PDG \cite{ParticleDataGroup:2026aaa} \\ \hline
$B^+ \to  K^{\ast +} {\bar K}^{\ast 0}$ \quad & \quad $0.55^{+0.14}_{-0.11}$ \quad  & \quad $0.06 \pm 0.01$ \quad & \quad $0.05 \pm 0.01$ \quad & \quad $0.66^{+0.15}_{-0.12}$  &$0.91\pm0.29$ \quad  \\
$B^+\to \rho^0\rho^+$ \qquad & \quad $13.5^{+4.0}_{-2.9}$ \quad & \quad $0.21^{+0.07}_{-0.05}$ \quad & \quad $0.22^{+0.07}_{-0.05}$ \quad & \quad $14.0^{+4.1}_{-3.1}$ &$24.0\pm1.9$ \quad \\ 
$B^+\to \rho^+ K^{\ast 0}$ \qquad & \quad $7.20^{+1.49}_{-1.18}$ \quad  & \quad $1.23^{+0.19}_{-0.18}$ \quad & \quad $0.97^{+0.15}_{-0.13}$ \quad & \quad $9.40^{+1.70}_{-1.64}$ \quad & \quad $9.2\pm1.5$ \quad \\
$B^+\to \rho^0 K^{\ast +}$ \qquad & \quad $4.98^{+1.09}_{-0.85}$ \quad & \quad $0.74^{+0.14}_{-0.11}$ \quad & \quad $0.53^{+0.08}_{-0.07}$ \quad & \quad $6.25^{+1.27}_{-1.00}$ \quad & \quad $4.6\pm1.1$ \quad \\ \hline     
$B^0 \to K^{\ast 0}\bar K^{\ast 0}$ \qquad & \quad $0.26^{+0.08}_{-0.05}$ \quad & \quad $0.07 \pm 0.01$ \quad & \quad $0.05 \pm 0.01$ \quad & \quad $0.38^{+0.09}_{-0.06}$ \quad & \quad $0.83\pm0.24$ \quad \\
{\bf LHCb} \cite{LHCb:2025ftm} & & & & & \quad $0.47\pm0.05$ \quad \\
$B^0\to K^{\ast +}K^{\ast -}$ \quad & \quad $0.17^{+0.05}_{-0.04}$ \quad & \quad $\sim 0$ \quad & \quad $\sim 0$ \quad & \quad $0.17^{+0.05}_{-0.04}$ \quad & \quad  $< 2.0 \quad $\\ 
$B^0\to \rho^+\rho^-$ \qquad & \quad $21.3^{+6.0}_{-4.6}$ \quad & \quad $0.74^{+0.21}_{-0.15}$ \quad & \quad $0.66^{+0.18}_{-0.13}$ \quad & \quad $22.7^{+6.3}_{-4.8}$ \quad & \quad $27.7\pm1.9$  \quad \\
{\bf Belle-II} \cite{Belle-II:2024frs} & & & & & \quad $28.9^{+0.37}_{-0.35}$ \quad \\
$B^0 \to \rho^0\rho^0$ \qquad & \quad $0.44^{+0.14}_{-0.10}$  \quad & \quad $0.05^{+0.02}_{-0.01}$ \quad & \quad $0.05 \pm 0.01$ \quad & \quad $0.54^{+0.16}_{-0.12}$ \quad & \quad $0.96\pm0.15$ \quad \\    
$B^0 \to \rho^-K^{\ast +}$ \qquad & \quad $6.60^{+1.33}_{-1.05}$ \quad & \quad $1.16^{+0.18}_{-0.17}$ \quad & \quad $0.96^{+0.14}_{-0.11}$ \quad & \quad $8.72^{+1.60}_{-1.30}$ \quad & \quad $10.3\pm2.6$ \quad \\  
$B^0 \to \rho^0 K^{\ast 0}$ \qquad & \quad $2.39^{+0.45}_{-0.37}$ \quad & \quad $0.54^{+0.09}_{-0.08}$ \quad & \quad $0.44\pm0.05$ \quad & \quad $3.37^{+0.57}_{-0.49}$ \quad & \quad $3.9\pm1.3$ \quad \\ \hline  
$B_s^0 \to  K^{\ast 0}\bar K^{\ast 0}$ \qquad & \quad $3.73^{+1.19}_{-0.85}$ \quad & \quad $1.33^{+0.22}_{-0.19}$ & \quad $1.11^{+0.18}_{-0.15}$ \quad & \quad $6.17^{+1.44}_{-1.06}$ \quad & \quad $11.1\pm2.7 $ \quad  \\ 
{\bf LHCb} \cite{LHCb:2025ftm} &  &  &  &  &  \quad $9.38\pm0.48$ \quad \\ 
$B_s^0 \to K^{\ast +}K^{\ast -}$ \qquad & \quad $5.55^{+1.80}_{-1.33}$ \quad & \quad $1.06^{+0.17}_{-0.13}$ \quad & \quad $0.94^{+0.15}_{-0.12}$ \quad & \quad $7.55^{+2.01}_{-1.50}$ \quad & \quad $---$ \\ 
$B_s^0 \to \rho^+\rho^-$ \qquad & \quad $1.47^{+0.25}_{-0.21}$ \quad & \quad $\sim 0$ \quad & \quad $\sim 0$ \quad & \quad $1.47^{+0.25}_{-0.21}$ \quad  & \quad $---$ \\   
$B_s^0 \to \rho^0\rho^0$ \qquad & \quad $0.72^{+0.11}_{-0.10}$ \quad & \quad $\sim 0$ \quad & \quad $\sim 0$ \quad & \quad $0.72^{+0.11}_{-0.10}$ \quad & \quad $<320$ \quad  \\   
$B_s^0 \to K^{\ast -}\rho^+$ \qquad & \quad $13.9^{+4.40}_{-3.49}$ \quad & \quad $0.44^{+0.14}_{-0.11}$ \quad & \quad $0.40^{+0.13}_{-0.10}$ \quad & \quad $14.7^{+4.68}_{-3.67}$  \quad & \quad $---$ \\
$B_s^0 \to \bar K^{\ast 0}\rho^0$ \qquad & \quad $0.62^{+0.21}_{-0.17}$ \quad & \quad $0.05^{+0.01}_{-0.02}$ \quad & \quad $0.04\pm0.01$ \quad & \quad $0.71^{+0.22}_{-0.20}$ \quad & \quad $<767$ \quad \\    
\hline \end{tabular} \end{center}
\end{table}

\begin{sidewaystable}[th] \begin{center}
\caption{Anatomy of the polarization fractions (in unit of $10^{-2}$) from each helicity amplitude.} \label{tab:polfrac} 
\setlength{\tabcolsep}{2pt}
\renewcommand{\arraystretch}{1.0}
\begin{tabular}{l  c c c  c c c  c c c}\hline 
Modes \quad & \quad $f_0^+$ \quad & \quad $f_0^-$ \quad & \quad $f_0$ \quad & \quad $f_{\parallel}^+$ \quad & \quad $f_{\parallel}^-$ \quad & \quad $f_{\parallel}$ \quad & \quad $f_{\perp}^+$ \quad & \quad $f_{\perp}^-$ \quad & \quad $f_{\perp}$ \quad \\ \hline
$B^+ \to  K^{\ast +} {\bar K}^{\ast 0}$  &$80.3^{+1.6}_{-1.9}$ &$86.0^{+1.2}_{-1.4}$ & $82.4^{+1.1}_{-1.1}$ &$13.2^{+1.5}_{-1.1}$ &$4.45^{+0.43}_{-0.35}$ &$10.0^{+0.8}_{-0.7}$  &$6.44^{+0.47}_{-0.37}$ &$9.52^{+1.09}_{-0.98}$  &$7.56^{+0.62}_{-0.53}$   \\
$B^+\to \rho^0\rho^+$  &$96.9^{+0.2}_{-0.1}$  &$96.9\pm0.1$  &$96.9\pm0.1$ &$1.50^{+0.03}_{-0.04}$&$1.50^{+0.03}_{-0.04}$ &$1.49^{+0.04}_{-0.02}$  &$1.56^{+0.05}_{-0.04}$&$1.56^{+0.05}_{-0.04}$  & $1.56^{+0.05}_{-0.04}$   \\ 
$B^+\to \rho^+ K^{\ast 0}$  &$76.9^{+1.6}_{-1.4}$ &$76.3\pm1.4$ &$76.6^{+1.5}_{-1.4}$  &$12.7\pm0.8$ &$13.4^{+0.7}_{-0.8}$ &$13.0\pm0.8$   &$10.4^{+0.6}_{-0.8}$ &$10.3^{+0.5}_{-0.7}$  &$10.3^{+0.6}_{-0.7}$    \\
$B^+\to \rho^0 K^{\ast +}$  &$66.3^{+2.5}_{-2.1}$  &$86.7\pm1.1$ &$80.0^{+1.1}_{-2.0}$ &$19.5\pm1.3$ &$7.7^{+0.8}_{-0.6}$  &$11.8\pm0.9$  &$14.1^{+1.0}_{-1.1}$ &$5.53^{+0.47}_{-0.45}$   &$8.51^{+0.65}_{-0.66}$    \\
{\bf LHCb} \cite{LHCb:2025zvw} \quad & \quad $49.1\pm8.7$ \quad & \quad $79.4\pm2.6$ \quad & \quad $72.0\pm2.9$ \quad & & & & & &   \\ 
{\bf HFLAV} \cite{HFLAV:2024ctg} \quad & & & \quad $78 \pm 12$ \quad & & & & & &   \\ \hline 
$B^0 \to K^{\ast 0}\bar K^{\ast0}$ \quad & \quad $69.0^{+3.9}_{-3.8}$ \quad & \quad $68.4^{+4.9}_{-5.0}$ \quad & \quad $68.8\pm5.3$ \quad & \quad $17.0^{+2.0}_{-2.1}$ \quad & \quad $17.7^{+2.3}_{-2.7}$ \quad & \quad $17.3\pm2.3$ \quad & \quad $14.0\pm1.8$ \quad & \quad $14.0^{+2.2}_{-2.3}$ \quad & \quad $14.0\pm2.0$ \\
{\bf LHCb} \cite{LHCb:2025ftm} \quad & & & \quad $60.0 \pm 2.8$ \quad & & & & & &      \\
{\bf HFLAV} \cite{HFLAV:2024ctg} \quad & & & \quad $73 \pm 5$ \quad & & & & & &      \\
$B^0\to K^{\ast +}K^{\ast -}$ \quad & \quad $\sim 100$ \quad & \quad $\sim 100$ \quad & \quad $\sim 100$ \quad & \quad $\sim 0$ \quad & \quad $\sim 0$ \quad & \quad $\sim 0$ \quad & \quad $\sim 0$ \quad & \quad $\sim 0$ \quad & \quad $\sim 0$ \quad    \\  
$B^0\to \rho^+\rho^-$ \quad & \quad $96.6^{+0.3}_{-0.2}$ \quad & \quad $90.9^{+0.4}_{-0.5}$ \quad & \quad $93.8\pm0.1$ \quad & \quad $1.81\pm0.13$ \quad & \quad $4.79^{+0.24}_{-0.20}$ \quad & \quad $3.27^{+0.06}_{-0.04}$ \quad & \quad $1.57\pm0.12$ \quad & \quad $4.30^{+0.22}_{-0.19}$ \quad & \quad $2.90^{+0.06}_{-0.02}$ \quad    \\     
{\bf Belle-II} \cite{Belle-II:2024frs} \quad & & & \quad $92.1 \pm 2.9$ \quad & & & & & &      \\
$B^0 \to \rho^0\rho^0$ \quad & \quad $82.3^{+2.1}_{-2.0}$ \quad & \quad $80.7\pm2.0$ \quad & \quad $80.9\pm1.9$ \quad & \quad $16.5^{+1.6}_{-1.9}$ \quad & \quad $9.96^{+0.76}_{-1.51}$ \quad & \quad $10.5\pm1.1$ \quad & \quad $1.21^{+0.41}_{-0.30}$ \quad & \quad $9.65^{+0.86}_{-0.93}$ \quad & \quad $8.56^{+0.89}_{-0.86}$ \quad   \\ 
$B^0 \to \rho^-K^{\ast +}$ \quad & \quad $64.8^{+2.1}_{-1.7}$ \quad & \quad $81.3^{+1.2}_{-1.3}$ \quad & \quad $75.7^{+1.6}_{-1.4}$ \quad & \quad $18.9^{+1.1}_{-1.2}$ \quad & \quad $10.3^{+0.7}_{-0.6}$ \quad & \quad $13.3^{+0.8}_{-0.9}$ \quad & \quad $16.2^{+0.8}_{-0.9}$ \quad & \quad $8.39^{+0.54}_{-0.55}$ \quad & \quad $11.0^{+0.6}_{-0.7}$ \quad   \\  
$B^0 \to \rho^0 K^{\ast0}$ \quad & \quad $72.3^{+1.6}_{-1.5}$ \quad & \quad $69.4\pm1.3$ \quad & \quad $71.0^{+1.5}_{-1.3}$ \quad & \quad $15.0\pm0.9$ \quad & \quad $17.3\pm0.6$ \quad & \quad $16.0\pm0.8$ \quad & \quad $12.7^{+0.6}_{-0.7}$ & \quad $13.3^{+0.6}_{-0.7}$ \quad & \quad $13.0^{+0.8}_{-0.7}$ \quad    \\ \hline  
$B_s^0 \to  K^{\ast 0} {\bar K}^{\ast 0}$ \quad & \quad $60.3^{+5.5}_{-5.4}$ \quad & \quad $60.5^{+5.3}_{-5.5}$ \quad & \quad $60.4\pm5.4$ \quad & \quad $21.6^{+2.9}_{-2.8}$ \quad & \quad $21.5^{+2.9}_{-2.8}$ \quad & \quad $21.6\pm2.9$ \quad & \quad $18.0^{+2.6}_{-2.5}$ \quad & \quad $18.0^{+2.5}_{-2.6}$ \quad & \quad $18.0^{+2.6}_{-2.5}$ \quad   \\  
{\bf LHCb} \cite{LHCb:2025ftm} \quad & & & \quad $15.9 \pm1.2$ \quad & & & & & &    \\  
$B_s^0 \to K^{\ast +}K^{\ast -}$ \quad & \quad $63.7^{+5.1}_{-5.5}$ \quad & \quad $81.6^{+3.2}_{-3.6}$ \quad & \quad $73.4^{+4.2}_{-4.5}$ \quad & \quad $19.2^{+2.8}_{-2.7}$ \quad & \quad $9.85^{+1.88}_{-1.67}$ \quad & \quad $14.1^{+2.3}_{-2.2}$ \quad & \quad $17.2^{+2.6}_{-2.5}$ \quad & \quad $8.58^{+1.72}_{-1.52}$ \quad & \quad $12.5^{+2.1}_{-2.0}$ \quad  \\   
$B_s^0 \to \rho^+\rho^- $ \quad & \quad $\sim100$ \quad & \quad $\sim100$ \quad & \quad $\sim100$ \quad & \quad $\sim 0$ \quad & \quad $\sim 0$ \quad & \quad $\sim 0$ \quad & \quad $\sim 0$ \quad & \quad $\sim 0$ \quad & \quad $\sim 0$ \quad  \\ 
$B_s^0 \to \rho^0\rho^0$ \quad & \quad $\sim100$ \quad & \quad $\sim100$ \quad & \quad $\sim100$ \quad & \quad $\sim 0$ \quad & \quad $\sim 0$ \quad & \quad $\sim 0$ \quad & \quad $\sim 0$ \quad & \quad $\sim 0$ \quad & \quad $\sim 0$ \quad   \\ 
$B_s^0 \to K^{\ast -}\rho^+$ \quad & \quad $97.4\pm0.3$ \quad & \quad $90.2^{+0.5}_{-0.6}$ \quad & \quad $94.3\pm0.1$ \quad & \quad $1.36\pm0.15$ \quad & \quad $5.11^{+0.31}_{-0.28}$ \quad & \quad $2.99^{+0.06}_{-0.05}$ \quad  & \quad $1.20\pm0.14$ \quad & \quad $4.71^{+0.30}_{-0.25}$ \quad & \quad $2.73^{+0.06}_{-0.05}$ \quad   \\  
$B_s^0 \to \bar K^{\ast 0} \rho^0$ \quad & \quad $90.8^{+0.4}_{-0.8}$ \quad & \quad $87.9^{+0.9}_{-1.1}$ \quad & \quad $88.2^{+0.8}_{-0.9}$ \quad & \quad $9.17^{+0.45}_{-0.59}$ \quad & \quad $5.85^{+0.62}_{-0.56}$ \quad & \quad $6.29^{+0.59}_{-0.56}$ \quad & \quad $0.0^{+0.36}_{-0.01}$ \quad & \quad $6.29^{+0.64}_{-0.49}$ \quad & \quad $5.47^{+0.62}_{-0.42}$ \quad  \\ \hline
\end{tabular}\end{center}  
\end{sidewaystable}

In table \ref{tab:BRs}, we present a detailed breakdown of the pQCD predictions for the branching fractions of $B_{(s)} \to \rho\rho, \rho K^\ast, K^\ast K^\ast$ decays. We compare our results with the world average values \cite{ParticleDataGroup:2026aaa}, as well as with recent measurements from the LHCb \cite{LHCb:2025ftm} and Belle-II \cite{Belle-II:2024frs} collaborations. The uncertainties in our calculations stem from the nonperturbative parameters of the mesons' LCDAs, and are dominated by the first inverse moment of the  $B$ meson \cite{Chai:2022ptk}. It is evident that the longitudinal helicity amplitude provides the dominant contribution across nearly all the decay modes considered, although its relative weight varies somewhat among these channels. Notably, in the penguin-dominated modes, namely $B \to \rho K^{\ast}$ and $B^0_{(s)} \to K^{\ast 0} {\bar K}^{\ast 0}, K^{\ast +} K^{\ast -}$, the longitudinal component is not expected to be overwhelmingly dominant. Instead, the perpendicular and parallel polarizations both contribute sizable to the total decay width. Overall, the pQCD predictions for the branching fractions are in good agreement with the experimental data. This is particularly evident for the penguin-dominated $B_{(s)} \to K^{\ast 0} {\bar K}^{\ast 0}$ decay, where the recent LHCb measurement with high statistics \cite{LHCb:2025ftm} yields a smaller central value compared to the world average. In addition, we reiterate the observation made in Ref.~\cite{Chai:2022ptk} that factorization-based predictions for the three $B \to \rho\rho$ channels obey the isospin relation, whereas the corresponding measurements exhibit a significant deviation. A similar discrepancy is also observed in the analogous $B \to \pi\pi$ channels.

The relative contributions of the different polarization amplitudes are clearly illustrated in Table \ref{tab:polfrac}, where the polarization fractions are dissected according to each helicity amplitude. The pQCD results are in good agreement with the available world averages \cite{HFLAV:2024ctg} for nearly all of the channels considered, consistent with our previous findings \cite{Chai:2022ptk}. Recently, the LHCb collaboration updated their measurements of the longitudinal polarization fractions for the decays $B^+ \to \rho^0 K^{\ast +}$ \cite{LHCb:2025zvw} and $B^0 \to K^{\ast 0} {\bar K}^{\ast 0}$ \cite{LHCb:2025ftm}. These new results, which benefit from significantly improved statistics, yield slightly smaller central values than the world averages while show better consistency with the pQCD predictions. However, a notable discrepancy is found for the decay $B_s \to K^{\ast 0} {\bar K}^{\ast 0}$, where the measured longitudinal polarization fraction is dramatically small. We highlight this as an intriguing polarization puzzle, as most theoretical frameworks, including QCD factorization (QCDF) \cite{Beneke:2006hg}, soft-collinear effective theory (SCET) \cite{Wang:2017rmh}, the factorization-assisted topological-amplitude approach (FAT) \cite{Wang:2017hxe}, as well as the pQCD approach  all favor a dominant longitudinal component. This persistent tension between theory and the latest LHCb measurement, shown in table \ref{tab:LPF}, thus poses a significant challenge to our understanding of penguin-dominated $B_s$ decays \cite{Beneke:2006hg,Huang:2005if,Li:2004mp}. Recently, the final-state interactions induced by rescattering have been proposed to explain the small longitudinal polarization fraction measured in $B_s \to K^{\ast 0} {\bar K}^{\ast 0}$ decay \cite{Yu:2024vlz}. For new physics investigations, see Refs. \cite{Choudhury:2026avj,Lizana:2023kei,Biswas:2023pyw}.

\begin{table}[b]\begin{center}
\caption{Comparison of theoretical predictions with recent LHCb measurements of the longitudinal polarization fraction in $B_s \to K^{\ast 0} {\bar K}^{\ast 0}$ decay and the associated U-spin symmetry ratio relative to the $B \to K^{\ast 0} {\bar K}^{\ast 0}$ channel.} \label{tab:LPF} 
\setlength{\tabcolsep}{6pt}
\renewcommand{\arraystretch}{1.2}
\begin{tabular}{ c c c c c c }\hline 
Observables \quad & \quad pQCD \quad &  \quad QCDF \cite{Beneke:2006hg} \quad & \quad SCET \cite{Wang:2017rmh} \quad & \quad FAT \cite{Wang:2017hxe} \quad & \quad LHCb \cite{LHCb:2025ftm} \quad \\ \hline
$f_0 \times 10^{2}$ \quad & \quad $60.4 \pm 5.4 $   \quad & \quad $63^{+42}_{-29}$ \quad  & \quad $44.9 \pm 18.3$ \quad & \quad $34.3 \pm 12.6$ \quad & \quad $15.9 \pm 1.2$ \quad  \\    
${\cal U}$ \quad & \quad $1.14^{+0.14}_{-0.13}$  \quad & \quad $1.10 ^{+1.07}_{-0.62}$ \quad  & \quad $1.11 ^{+0.85}_{-0.47}$ \quad & \quad $1.70 ^{+1.04}_{-0.56}$ \quad & \quad $3.77 ^{+0.36}_{-0.31}$ \quad  \\ \hline \end{tabular} \end{center}\vspace{-8mm}
\end{table}

\begin{sidewaystable}[th] \begin{center}
\caption{Anatomy of the CP asymmetries (in unit of $10^{-2}$) from each helicity amplitude.} \label{tab:cv} 
\setlength{\tabcolsep}{4pt}
\renewcommand{\arraystretch}{1.2}
\begin{tabular}{l  c c  c c  c c  c } \hline 
Modes \quad & \quad $\rm {\mathcal{A}_0^{CP}}$ \quad & \quad $\Delta \mathcal{A}_0^{CP}$ \quad & \quad $\rm {\mathcal{A}_{\parallel}^{CP}}$ \quad & \quad $\Delta \mathcal{A}_{\parallel}^{CP}$ \quad & \quad $\rm {\mathcal{A}_{\perp}^{CP}}$ \quad & \quad $\Delta \mathcal{A}_{\perp}^{CP}$ \quad & \quad $\rm {\mathcal{A}^{CP}}$ \quad \\   \hline
$B^+ \to K^{\ast +} {\bar K}^{\ast 0}$ \quad & \quad $-23.6^{+5.4}_{-4.8}$ \quad & \quad $0.76\pm0.01$ \quad & \quad $-67.5^{+2.6}_{-2.3}$ \quad & \quad $0.81^{+0.07}_{-0.06}$ \quad & \quad $-7.87^{+1.83}_{-1.75}$ \quad & \quad $0.97\pm0.02$ \quad & \quad $-26.8^{+2.5}_{-3.1}$ \quad  \\ 
$B^+\to \rho^0\rho^+$ \quad & \quad $0.04 \pm 0.01$ \quad & \quad $-2.43\pm 0.01$ \quad & \quad $0.04 \pm 0.01$ \quad & \quad $-2.43^{+0.01}_{-0.02}$ \quad & \quad $-0.03 \pm 0.01$ \quad & \quad $-2.43^{+0.0}_{-0.02}$ \quad & \quad $0.03 \pm 0.01$ \quad  \\ 
$B^+\to \rho^+ K^{\ast 0}$ \quad & \quad $0.15\pm0.28$ \quad & \quad $0.02 \pm 0.01$ \quad & \quad $3.22^{+0.27}_{-0.25}$ \quad & \quad $0.01 \pm 0.01$ \quad & \quad $0.50\pm0.06$ \quad & \quad $-0.01 \pm 0.01$ \quad & \quad $0.58^{+0.21}_{-0.22}$ \quad  \\
$B^+\to \rho^0 K^{\ast +}$ \quad & \quad $42.2^{+1.2}_{-1.5}$ \quad & \quad $0.44^{+0.04}_{-0.05}$ \quad & \quad $-14.6\pm1.0$ \quad & \quad $0.12\pm0.04$ \quad & \quad $-15.1^{+1.1}_{-0.8}$ \quad & \quad $0.20^{+0.04}_{-0.03}$  \quad & \quad $30.6^{+0.5}_{-0.7}$ \quad \\
{\bf LHCb} \cite{LHCb:2025zvw} \quad & \quad $66.4\pm8.4$ \quad & \quad $0.72\pm0.18$ \quad & \quad $-6.3\pm14.0$ \quad & \quad $0.48\pm0.20$ \quad & \quad $28.4\pm14.9$ \quad & \quad $0.41\pm0.22$ \quad & \quad $50.7\pm6.4$  \quad \\   \hline 
$B^0 \to K^{\ast 0}\bar K^{\ast 0}$ \quad & \quad $-17.9\pm0.4$ \quad & \quad $0.69 \pm 0.01$ \quad & \quad $-15.5^{+1.7}_{-1.9}$ \quad & \quad $0.71 \pm 0.01$ \quad & \quad $-17.6^{+2.0}_{-2.2}$ \quad & \quad $0.71 \pm 0.01$ \quad & \quad $C=-17.4^{+0.5}_{-0.4}$, $S=-4.46^{+0.79}_{-0.42}$  \\
$B^0\to K^{\ast +}K^{\ast -}$ \quad & \quad $-1.69^{+2.30}_{-5.40}$ \quad & \quad $-1.15^{+0.08}_{-0.13}$ \quad & \quad $36.1^{+1.0}_{-1.5}$ \quad & \quad $-2.16^{+0.05}_{-0.06}$ \quad & \quad $-17.1\pm0.1$ \quad & \quad $3.80^{+0.01}_{-0.0}$ \quad & \quad $C=-1.67^{+2.30}_{-5.40}$, $S=-94.9^{+4.8}_{-2.0}$ \quad  \\
$B^0\to \rho^+\rho^-$ \quad & \quad $-4.89^{+0.54}_{-0.49}$ \quad & \quad $-2.28 \pm 0.01$ \quad & \quad $43.7^{+4.7}_{-4.6}$ \quad & \quad $-2.19^{+0.01}_{-0.02}$ \quad & \quad $45.1^{+5.0}_{-4.8}$ \quad & \quad $-2.23^{+0.02}_{-0.01}$ \quad & \quad $C=-1.85^{+0.20}_{-0.11}$, $S=-12.7^{+0.4}_{-0.3}$ \quad \\
{\bf Belle-II} \cite{Belle-II:2024frs} \quad & \quad  \quad & \quad \quad & \quad \quad & \quad \quad & \quad \quad & \quad \quad & \quad   $C=-2.00 \pm 13.0$, $S=-26.0 \pm 20.0$ \quad \\   \hline 
$B^0 \to \rho^0\rho^0$ \quad & \quad $74.2^{+2.6}_{-3.3}$ \quad & \quad $-2.42^{+0.02}_{-0.04}$ \quad & \quad $60.3^{+2.9}_{-3.8}$ \quad & \quad $-1.08\pm0.10$ \quad & \quad $96.4^{+0.7}_{-1.3}$ \quad & \quad$-3.67^{+5.91}_{-0.09}$ \quad & \quad $C=74.6^{+2.3}_{-3.0}$, $S=1.38^{+2.27}_{-1.93}$ \quad \\
$B^0 \to \rho^-K^{\ast +}$ \quad & \quad $42.1^{+0.6}_{-1.1}$ \quad & \quad $0.56^{+0.04}_{-0.05}$ \quad & \quad $3.28^{+0.29}_{-0.26}$ \quad & \quad $-0.01^{+0.04}_{-0.01}$ \quad & \quad $0.53^{+0.06}_{-0.05}$ \quad & \quad $-0.01\pm0.01$ \quad & \quad $C=32.4^{+0.1}_{-0.2}$, $S=-16.2^{+4.6}_{-4.3}$ \quad  \\
$B^0 \to \rho^0 K^{\ast 0}$ \quad & \quad $-16.4^{+1.8}_{-2.0}$ \quad & \quad $0.19\pm0.02$ \quad & \quad $-7.38\pm0.72$ \quad & \quad $0.10^{+0.01}_{-0.02}$ \quad & \quad $-12.2^{+1.1}_{-1.2}$ \quad & \quad $0.01^{+0.01}_{-0.02}$ \quad & \quad $C=-14.4^{+1.5}_{-1.7}$, $S=-51.1^{+1.8}_{-2.2}$ \quad  \\ \hline   
$B_s^0 \to  K^{\ast 0} {\bar K}^{\ast 0}$ \quad & \quad $0.87\pm0.03$ \quad & \quad $0.01 \pm 0.01$ \quad & \quad $0.55^{+0.09}_{-0.08}$ \quad & \quad $0.00 \pm 0.01$ \quad & \quad $0.62^{+0.09}_{-0.10}$ \quad & \quad $0.00 \pm 0.01$ \quad & \quad $C=0.75\pm0.04$, $S=-66.8^{+0.0}_{-0.1}$ \quad \\
$B_s^0 \to K^{\ast +}K^{\ast -}$ \quad & \quad $21.3^{+2.4}_{-2.2}$ \quad & \quad $-5.64^{+0.0}_{-0.02}$ \quad & \quad $-23.5\pm1.8$ \quad & \quad $0.20^{+0.03}_{-0.02}$ \quad & \quad $-24.9\pm2.0$ \quad & \quad $0.19\pm0.03$ \quad & \quad $C=9.24^{+1.67}_{-1.73}$, $S=-9.91^{+0.54}_{-0.94}$ \quad \\
$B_s^0 \to \rho^+\rho^-$ \quad & \quad $2.87^{+0.59}_{-0.82}$ \quad & \quad $0.14\pm0.0$ \quad & \quad $-32.4\pm0.2$ \quad & \quad $0.08\pm0.01$ \quad & \quad $55.7\pm0.46$ \quad & \quad $0.67\pm0.0$ \quad & \quad $C=2.86^{+0.60}_{-0.81}$, $S=-56.2^{+0.4}_{-0.2}$ \quad \\
$B_s^0 \to \rho^0\rho^0$ \quad & \quad $2.36^{+0.47}_{-0.56}$ \quad & \quad $0.15\pm0.0$ \quad & \quad $-30.6\pm0.2$ \quad & \quad $0.06\pm0.01$ \quad & \quad $62.3^{+0.5}_{-0.3}$ \quad & \quad $0.59^{+0.02}_{-0.01}$ \quad & \quad $C=2.36^{+0.45}_{-0.56}$, $S=-55.4^{+0.1}_{-0.3}$ \quad \\
$B_s^0 \to K^{\ast -}\rho^+$ \quad & \quad $-16.7^{+1.5}_{-1.6}$ \quad & \quad $-2.19 \pm 0.01$ \quad & \quad $48.6^{+5.3}_{-5.2}$ \quad & \quad $-2.08^{+0.02}_{-0.04}$ \quad & \quad $50.3^{+5.6}_{-5.4}$ \quad & \quad $-2.11^{+0.02}_{-0.04}$ \quad & \quad $C=-12.9^{+1.1}_{-1.2}$, $S=-20.7^{+0.8}_{-1.7}$ \quad \\
$B_s^0 \to \bar K^{\ast 0} \rho^0$ \quad & \quad $72.9^{+3.4}_{-3.1}$ \quad & \quad $-1.73^{+0.08}_{-0.07}$ \quad & \quad $61.5^{+5.2}_{-4.5}$ \quad & \quad $-1.0^{+0.20}_{-0.18}$ \quad & \quad $99.7^{+0.2}_{-1.7}$ \quad & \quad $4.19^{+0.69}_{-4.16}$ \quad & \quad $C=73.6^{+3.4}_{-3.1}$, $S=-42.8^{+5.4}_{-5.2}$ \quad \\ \hline
\end{tabular}\end{center}  
\end{sidewaystable}

The decays $B \to K^{\ast 0} {\bar K}^{\ast 0}$ and $B_s \to K^{\ast 0} {\bar K}^{\ast 0}$ share nearly identical penguin-dominated topological structures, with all amplitudes first entering at NLO. Their difference stems primarily from the corresponding CKM matrix elements, as detailed in the explicit decay amplitude expressions given in Refs. \cite{Yan:2018fif,Chai:2022ptk}. To mitigate common systematic uncertainties, we focus on the U-spin ratio of the longitudinal polarization fractions,
\beq {\cal U} = \frac{f_0(B^0 \to K^{\ast 0} {\bar K}^{\ast 0})}{f_0({B_s^0 \to K^{\ast 0} {\bar K}^{\ast 0}})}.
\label{eq:UR}\eeq
With modest U-spin breaking of order $\mathcal{O}(m_s/\Lambda_{\rm QCD})$, this ratio is expected to be close to unity, a result supported by various theoretical approaches, including the pQCD method adopted here. In stark contrast, the recent LHCb measurement \cite{LHCb:2025ftm} deviates from unity by more than six standard deviations. If confirmed by independent measurements, this discrepancy would signal either large non-factorizable spectator contributions, or otherwise point to new physics beyond the Standard Model.

Table \ref{tab:cv} presents the pQCD results for the helicity-dependent CP asymmetries of the squared amplitudes and their relative phases. Within uncertainties, the predicted direct CP asymmetries are consistent with the world average, as what we have seem in Ref. \cite{Chai:2022ptk}. When compared to the first measurement of helicity-dependent CP asymmetries in $B^+ \to \rho^0 K^{\ast +}$ decay \cite{LHCb:2025zvw}, the pQCD calculation confirms the longitudinally dominated mechanism and reproduces the relatively large CP asymmetries, although a discrepancy of $3 \sigma$ to $4 \sigma$ remains. In particular, the predicted transverse CP asymmetry for the squared magnitude has the opposite sign to the data.

\section{Summary}~\label{sec:summary}

We perform a pQCD analysis of $B_{(s)} \to \rho\rho, \rho K^\ast$ and $K^\ast K^\ast$ decays, incorporating all known NLO corrections and focusing on polarization observables and helicity-dependent CP asymmetries. Our predictions for branching fractions and longitudinal polarization fractions agree well with experimental data for most channels, particularly for $B^0 \to K^{\ast 0} {\bar K}^{\ast 0}$ and $B^+ \to \rho^0 K^{\ast +}$. However, two significant discrepancies emerge: (i) the predicted longitudinal polarization fraction $B_s \to K^{\ast 0} {\bar K}^{\ast 0}$, $\left( 60.4 \pm 5.4 \right) \%$, far exceeds the LHCb measurement, $\left( 15.9 \pm 1.2 \right) \%$, yielding a U-spin ratio ${\cal U} = 1.14^{+0.14}_{-0.13}$ versus the observed $3.77^{+0.36}_{-0.31}$; (ii) the predicted direct CP asymmetry in $B^+ \to \rho^0 K^{\ast +}$ is about $40 \%$ smaller than the measured value, with the transverse component showing the opposite sign. These tensions highlight the need for improved theoretical treatments of penguin-dominated $B_{(s)}$ decays and may indicate either large spectator contributions at NLO or physics beyond the Standard Model.

\section{ACKNOWLEDGEMENTS}

We are grateful to Yue-Hong Xie, Hang Yin, and Yan-Xi Zhang for valuable discussions. This work was supported by the National Natural Science Foundation of China (NSFC) under Grant No. 12575098, 12505107, 12675119, the Hunan Province Major Basic Research Projects under Grant No. 2026JC0006 (S.C.), and the Launching Funding of Henan University of Technology under Grant No. 2024BS052 (J.C.).


\begin{thebibliography}{99}

\bibitem{Gambino:2020jvv}
P.~Gambino, A.~S.~Kronfeld, M.~Rotondo, C.~Schwanda, F.~Bernlochner, A.~Bharucha, C.~Bozzi, M.~Calvi, L.~Cao and G.~Ciezarek, \textit{et al.}
Eur. Phys. J. C \textbf{80}, no.10, 966 (2020)

\bibitem{Cerri:2018ypt}
A.~Cerri \textit{et al.,}  
CERN Yellow Rep. Monogr. \textbf{7}, 867-1158 (2019).

\bibitem{Kou:2018nap}
E.~Kou \textit{et al.} [Belle-II], PTEP \textbf{2019}, 123C01 (2019) [erratum: PTEP \textbf{2020}, 029201 (2020)].

\bibitem{Korner:1979ci}
J.~G.~Korner and G.~R.~Goldstein,
Phys. Lett. B \textbf{89}, 105-110 (1979)

\bibitem{Wirbel:1985ji}
M.~Wirbel, B.~Stech and M.~Bauer,
Z. Phys. C \textbf{29}, 637 (1985)

\bibitem{Bauer:1986bm}
M.~Bauer, B.~Stech and M.~Wirbel,
Z. Phys. C \textbf{34}, 103 (1987)

\bibitem{Ali:2007ff}
A.~Ali, G.~Kramer, Y.~Li, C.~D.~Lu, Y.~L.~Shen, W.~Wang and Y.~M.~Wang,
Phys. Rev. D \textbf{76} (2007), 074018.

\bibitem{Chen:2002pz}
C.~H.~Chen, Y.~Y.~Keum and H.~n.~Li,
Phys. Rev. D \textbf{66}, 054013 (2002)

\bibitem{Li:2004ti}
H.~n.~Li and S.~Mishima,
Phys. Rev. D \textbf{71}, 054025 (2005)

\bibitem{Zou:2015iwa}
Z.~T.~Zou, A.~Ali, C.~D.~Lu, X.~Liu and Y.~Li,
Phys. Rev. D \textbf{91} (2015), 054033.

\bibitem{Yan:2018fif}
D.~C.~Yan, X.~Liu and Z.~J.~Xiao,
Nucl. Phys. B \textbf{935} (2018), 17-39.

\bibitem{Beneke:2006hg}
M.~Beneke, J.~Rohrer and D.~Yang,
Nucl. Phys. B \textbf{774}, 64-101 (2007)

\bibitem{Cheng:2009cn}
H.~Y.~Cheng and C.~K.~Chua,
Phys. Rev. D \textbf{80}, 114008 (2009)

\bibitem{Wang:2017rmh}
C.~Wang, S.~H.~Zhou, Y.~Li and C.~D.~Lu,
Phys. Rev. D \textbf{96}, no.7, 073004 (2017)

\bibitem{Yan:2025ocu}
D.~C.~Yan, H.~n.~Li, Z.~Rui, Z.~J.~Xiao and Y.~Li,
Eur. Phys. J. C \textbf{85}, no.4, 444 (2025)

\bibitem{Li:2021qiw}
Y.~Li, D.~C.~Yan, Z.~Rui and Z.~J.~Xiao,
Eur. Phys. J. C \textbf{81}, no.9, 806 (2021)

\bibitem{Colangelo:2004rd}
P.~Colangelo, F.~De Fazio and T.~N.~Pham,
Phys. Lett. B \textbf{597}, 291-298 (2004)

\bibitem{Cheng:2004ru}
H.~Y.~Cheng, C.~K.~Chua and A.~Soni,
Phys. Rev. D \textbf{71}, 014030 (2005)

\bibitem{LHCb:2025zvw}
R.~Aaij \textit{et al.} [LHCb],
Phys. Rev. Lett. \textbf{136} (2026) no.2, 021803

\bibitem{Belle-II:2024frs}
I.~Adachi \textit{et al.} [Belle-II],
Phys. Rev. D \textbf{111}, no.9, 092001 (2025)

\bibitem{LHCb:2025ftm}
R.~Aaij \textit{et al.} [LHCb],
Phys. Rev. D \textbf{113} (2026) no.9, 092002

\bibitem{Choudhury:2026avj}
D.~Choudhury, S.~Kumbhakar, A.~Kundu and S.~Nandi,
Phys. Rev. D \textbf{114}, no.1, 015019 (2026)

\bibitem{Chai:2022ptk}
J.~Chai, S.~Cheng, Y.~h.~Ju, D.~C.~Yan, C.~D.~L{\"u} and Z.~J.~Xiao,
Chin. Phys. C \textbf{46} (2022) no.12, 123103.

\bibitem{Li:2012nk}
H.~n.~Li, Y.~L.~Shen and Y.~M.~Wang,
Phys. Rev. D \textbf{85}, 074004 (2012)

\bibitem{Cheng:2014fwa}
S.~Cheng, Y.~Y.~Fan, X.~Yu, C.~D.~L{\"u} and Z.~J.~Xiao,
Phys. Rev. D \textbf{89}, no.9, 094004 (2014)

\bibitem{Beneke:2000ry}
M.~Beneke, G.~Buchalla, M.~Neubert and C.~T.~Sachrajda,
Nucl. Phys. B \textbf{591}, 313-418 (2000)

\bibitem{Beneke:2001ev}
M.~Beneke, G.~Buchalla, M.~Neubert and C.~T.~Sachrajda,
Nucl. Phys. B \textbf{606}, 245-321 (2001)

\bibitem{Beneke:2003zv}
M.~Beneke and M.~Neubert,
Nucl. Phys. B \textbf{675}, 333-415 (2003)

\bibitem{Liu:2015sra}
X.~Liu, H.~n.~Li and Z.~J.~Xiao,
Phys. Rev. D \textbf{91}, no.11, 114019 (2015)

\bibitem{ParticleDataGroup:2026aaa}
F.~Takahashi \textit{et al.} [Particle Data Group],
Int. J. Mod. Phys. A 41, 2630011 (2026).

\bibitem{HFLAV:2024ctg}
S.~Banerjee \textit{et al.} [Heavy Flavor Averaging Group (HFLAV)],
Phys. Rev. D \textbf{113}, no.1, 012008 (2026)

\bibitem{Wang:2017hxe}
C.~Wang, Q.~A.~Zhang, Y.~Li and C.~D.~Lu,
Eur. Phys. J. C \textbf{77} (2017) no.5, 333.

\bibitem{Huang:2005if}
H.~W.~Huang, C.~D.~Lu, T.~Morii, Y.~L.~Shen, G.~Song and Jin-Zhu,
Phys. Rev. D \textbf{73}, 014011 (2006)

\bibitem{Li:2004mp}
H.~n.~Li,
Phys. Lett. B \textbf{622}, 63-68 (2005)

\bibitem{Yu:2024vlz}
Y.~Yu, H.~B.~Fu, H.~Zhang and B.~C.~Ke,
Eur. Phys. J. C \textbf{85}, no.1, 42 (2025)

\bibitem{Lizana:2023kei}
J.~M.~Lizana, J.~Matias and B.~A.~Stefanek,
JHEP \textbf{09}, 114 (2023)

\bibitem{Biswas:2023pyw}
A.~Biswas, S.~Descotes-Genon, J.~Matias and G.~Tetlalmatzi-Xolocotzi,
JHEP \textbf{06}, 108 (2023)

\end{thebibliography}
\end{document}